\documentstyle[prl,aps]{revtex} 

\begin{document}

\draft


\title{RELATIVISTIC SYMMETRY AND ENTANGLED CORRELATIONS}


\author{Michael E. Kellman}

\address{Institute of Theoretical Science,
\linebreak University of Oregon, Eugene, OR 97403 \linebreak
kellman@oregon.uoregon.edu}

\date{\today}

\maketitle

\begin{abstract}

It is argued that the standard quantum mechanical description of the Bell correlations between entangled subsystems is in conflict with relativistic space-time symmetry.   Proposals to abandon relativistic symmetry, in the sense of explicitly returning to an absolute time and preferred frame, are rejected on the grounds that the preferred frame is not empirically detectable, so the asymmetry is an unsatisfactory feature in physical theory.  A ``symmetric view" is proposed in which measurement  events on space-like separated entangled subsystems are connected by a symmetric two-way mutual influence.  Because of this reciprocity, there is complete symmetry of the description:  Einsteinian relativity of simultaneity and space-time symmetry are completely preserved.  The nature of the two-way influence is considered, as well as the possibility of an empirical test.  

\end{abstract}





\date{\today}

\maketitle





\section{Introduction} \label{introduction}

Bell's theorem \cite{Bellstheorem,speakable} shows that by the rules of quantum mechanics, entangled systems have nonseparable correlations between measurements performed on subsystems which are space-like separated.    One way of putting it is that the outcomes of measurements are not pre-determined by any ``real state" of the subsystems existing independently of the measurements.  Experiments on EPR-Bohm systems \cite{epr,bohmepr} uphold the predictions of quantum mechanics, and while there is some room for the possibility of loopholes, it is hard not to conclude that there is some kind of information transfer taking place over space-like separated intervals.   Since it is not possible to send a signal by means of the Bell correlations, it is useful to speak of an ``influence" which apparently travels at faster-than-light velocities, in fact, by the experimental evidence, ``instantaneously".  

If this is so, it appears to be in conflict with special relativity, in the Einsteinian sense of the relativity of space and time (as opposed to the earlier Lorentzian view in which relativistic phenomena occur as dynamical effects against a background of absolute space and time).  The problem is not the faster-than-light nature of the influence {\it per se}.  Rather, it is the grave problems this seems to present for causality.  In standard quantum mechanical description, measurement on a subsystem 1 of an entangled system exerts an influence on the other subsystem 2, resulting in correlation in the outcome of a subsequent measurement on 2.  However, viewed from another reference frame, according to relativity it is the measurement on 2 which is temporally prior, and exerts an influence on 1.  (See Fig.~\ref{standardview}.) The quantum mechanical description of the apparent causal sequence -- not only the time ordering, but what is cause and what is effect -- thus has an asymmetry with respect to choice of reference frame.  This seems a gross violation of the idea of Einsteinian relativity.    

Some workers \cite{Bellghost,Bohmghost,Hileyghost,BellreBohmspeak,undivided}, including Bell himself, take this  conflict with relativity very seriously, and are willing to consider a drastic response.  They propose to abandon the relativity of simultaneity and return to the idea of absolute space and time as the arena, perhaps embedding a sort of quantum mechanical ``ether", in which quantum events take place.  The special relativity view of space-time is then regarded as holding only for an ensemble of systems, a kind of statistical mirage as it were.  Bell points out \cite{Bellspeakoldrel,Bellghostoldrel} that such a viewpoint is completely consistent with the facts of classical relativistic mechanics.  In fact, the notion of absolute space-time and a classical ether lay at the heart of the pre-Einsteinian relativity of Lorentz.  Furthermore, an absolute reference frame is certainly compatible with the known facts about quantum mechanical systems.   

On the other hand, there are good reasons to resist the notion of the incompatibility of Einsteinian relativity and quantum phenomena, and to abhor the suggestion of a return to an absolute reference frame.    The statistical predictions on ensembles of systems, including the correlations between entangled subsystems, are the same regardless of the choice of  frame.   This implies that within the currently understood quantum mechanical theory, there is no empirical means whatsoever to detect the supposed absolute frame.  It seems that nature takes great care to mask the apparent inconsistency between relativity and quantum mechanics.   The situation appears much like that which faced Einstein when he developed special relativity.  As Einstein put it \cite{EinsteinSR05} regarding Maxwell's electrodynamics, the theory ``leads to asymmetries which do not appear to be inherent in the phenomena".  This seemed unsatisfactory, even though there was no conflict of the theory with experiment.  Exactly the same statement pertains to the present situation.  It is hard not to think that nature is trying to tell us something very interesting with this state of affairs. 

The view proposed here is that one should maintain relativistic symmetry, including the complete equivalence and indistinguishability of reference frames.  It is neither necessary nor desirable to return to the idea of absolute space and time.  Rather, the nonrelativistic features of the quantum formalism should be remedied with a change in the description of quantum phenomena, including revision of notions of causality -- though not necessarily of the predictions of quantum mechanics.  

Consistency with the quantum mechanical facts necessitates a faster-than-light quantum influence between measurements on entangled sub-systems which are space-like separated.  The proposal here is that this is a reciprocal,  two-way influence:  there is a {\it mutuality of causation} which is entirely symmetrical with respect to reference frame.  This makes it possible to remove the apparent conflict with space-time symmetry in the standard description.   

It should be noted that Shimony has briefly but suggestively \cite{Shimonynewphysics} used language similar to this in describing the uneasy ``peaceful coexistence" of relativity and quantum mechanics.  The formulation and development of the ideas presented here reflects the thinking of the present author.      

I present a further description of the conflict of relativity and quantum mechanics in Section \ref{theproblem}.  I outline several solutions proposed by others, and what I see as objections to those proposals in Section \ref{others}.  The reader who is already convinced that there is a conflict may wish to skip Section \ref{theproblem}; the reader already persuaded that none of the earlier proposed solutions is adequate to the task may wish to go directly to the proposal in Section \ref{symview}.  

\section{Relativistic asymmetry of current quantum mechanical description} \label{theproblem}

This section discusses the conflict between relativity and quantum mechanical description as viewed through the lenses of several different ``standard" interpretations.  

Initially the wave function is entangled, i.e. a nonseparable sum of products

\begin{equation} |\psi_0 \rangle = | z_+(1)\rangle  |z_-(2)\rangle  - |z_-(1)\rangle |z_+(2)\rangle  \end{equation}

\noindent with the notation indicating spin + along the $z$ direction for particle 1, etc.  Then the {\it a priori} probability of finding the outcome $ \mu_i$ where $i = \pm$ when measuring spin in the $ \mu$ direction on 1 is 

\begin{equation} P(\mu_i (1)) = |\langle \mu_i | z_+ \rangle|^2  +  |\langle \mu_i | z_- \rangle|^2  \end{equation}

\noindent and the {\it a priori} probability of finding the outcome $ \nu_i$ when measuring spin in the $ \nu$ direction on 2 is 

\begin{equation} P(\nu_i (2)) = |\langle \nu_i | z_- \rangle|^2  +  |\langle \nu_i | z_+ \rangle|^2.  \end{equation}

\noindent  However, according to the standard quantum mechanical calculation, if spin in direction $\mu$ is measured on 1 and spin in direction $\nu$ on 2, the probability of finding the joint outcome $\mu_i(1) \nu_j(2)$ is not given by the product of the individual {\it a priori} probabilities:

\begin{equation}  P (\mu_i(1) \nu_j(2))  \ne  P(\mu_i (1)) P(\nu_i (2)). \label{inequality} \end{equation}

\noindent  The inequality (\ref{inequality}) is basically the statement that something like the Bell correlations exist.  If the measurement on 2 is later than the prior measurement on 1, then the probability of the outcome $\nu_i(2)$ has been conditioned by the prior measurement on 1 to have a value different from $P(\nu_i (2))$.  It seems reasonable to say that the earlier measurement on 1 has ``influenced" the later measurement on 2, i.e. influenced the probabilities of its possible outcomes.    This leads to the conclusion that the outcomes of measurements are not pre-determined by any ``real state" of the subsystems existing independently of the measurements; an influence of the measurement process is necessary to account for the empirical observations.

On the other hand, there are frames moving relative to the frame just described, in which the measurement on 1 is later than the measurement on 2.  It then seems reasonable in these frames to say that it is the probability of the outcome $\mu_i(1)$ which is conditioned by the prior measurement on 2 to have a value different from $P(\mu_i (1))$.  It is reasonable to conclude in these frames that it is the earlier measurement on 2 which has influenced the later measurement on 1.  

The problem of relativistic noninvariance is evident.  In one reference frame measurement on unconditioned particle 1 influences the outcome of  later measurement on 2; while in the second frame, the same ensemble of events  is described in terms of measurement on unconditioned particle 2, which influences later measurement on 1.     

These statements about the existence of a quantum mechanical influence which is not relativistically invariant seem equally applicable to any of various ``standard" interpretations of quantum mechanics, whether one posits a collapse of the wave function, or a ``many-worlds" interpretation, or a de Broglie-Bohm ``hidden variables" picture.  

The problem of relativistic noninvariance appears with special vividness in the latter picture, and for this reason, it is worth considering from this point of view how the Bell correlations come about.  Like many-worlds, de Broglie-Bohm is a no-collapse description in which the measurement apparata as well as the system are included in the quantum mechanical state.   However, unlike many-worlds there is no ``splitting".    

Bohm and Hiley \cite{undivided} describe in detail the measurement process for the EPR-Bohm experiment.  The picture goes like this.  Measurement on particle 1  in the $\mu$ direction gives a result predetermined by the ``hidden variables" of the particle and the measurement apparatus.  The measurement results in a change of the wave function and the nonlocal ``quantum potential".  This causes a very rapid, essentially instantaneous ``swerve" in the variables of the other particle 2. Then subsequent measurement on particle 2  gives results which are correlated with the measurement on particle 1, because of the swerve in the variables of particle 1.  All of this is described by the time development of the wave function for the system plus measurement devices.    

This description is manifestly nonrelativistic.  In another frame, the first measurement is on particle 2, causing particle 1 to swerve, according to the instantaneous change in the quantum potential, so in this frame a different measurement event causes a different particle to swerve.  Furthermore, the outcomes for the {\it individual} particles in general are frame-dependent!  Consider a measurement in the $\mu$ direction for both particles.  Suppose in the frame where measurement on 1 takes place first, the deterministic evolution gives spin + for particle 1, then spin $-$ for particle 2.  Then in the frame where measurement at B takes place first, the result is spin + for particle 2, then spin $-$ for particle 1 \cite{Albert}.  

For some, the acuteness with which the problem of relativistic asymmetry appears in the de Broglie-Bohm interpretation is good cause for its rejection.  Bell however was of the opinion \cite{Bellsix} that the problem with relativity is not  worse in the de Broglie-Bohm picture, only that it is brought more clearly to the fore.  I find Bell's view persuasive;  the reason for considering the de Broglie-Bohm picture here is to highlight just how much trouble there is when one tries to bring the standard interpretations of quantum mechanics face-to-face with relativity.   

In the quantum mechanical description, the conflict with relativity comes as an inconsistency in the causal sequence of events that take place when viewed in different reference frames.  It is not surprising that the effort to merge relativity and the quantum mechanics of entangled systems would lead to a problem with causality. It was of course well-understood in classical relativity that a conflict with traditional ideas of causality is avoided only because faster-than-light signals or influences are excluded.   The present conflict is the inevitable result of trying to have three things:  traditional  notions of causality, a faster-than-light influence in the quantum mechanics of entangled systems, and relativistic symmetry with relativity of simultaneity.  Since the experimental facts seem to demand the existence of a faster-than-light influence, it appears that either relativistic symmetry or traditional causality must be abandoned or modified.

\section{proposed solutions} \label{others}

I consider briefly several of the ideas proposed by others as solutions to the conflict of relativity and quantum mechanics, and why in the end I find each of them unsatisfactory.  These proposals fall into two broad classes:  views which basically accept the conflict, and try to deal with it in a way which violates relativistic symmetry; and views which try to retain the relativistic symmetry, at the cost of a drastic change in ideas of causality.

\subsection {quantum mechanical ether or preferred frame}

Some, including Bell \cite{Bellghost,BellreBohmspeak}, Bohm \cite{Bohmghost,undivided}, and Hiley \cite{Hileyghost,undivided}, are willing to consider a return to the idea of a universal or absolute preferred frame, with absolute time; Bell associates this with a kind of quantum mechanical ether \cite{realcausseq}.  This seems to be in the context of a preference for the de Broglie-Bohm interpretation as that which comes closest among ``standard" interpretations to satisfying Bell's expressed desire \cite{realcausseq} for a description of the Bell correlations that involves a ``real causal sequence".  However, I find the idea of a preferred frame unsatisfactory for the reasons discussed in the Introduction.  It should be said that ideas of an absolute frame or quantum mechanical ether are probably not the last word of what either Bell or Bohm  hoped for in some future understanding of quantum phenomena \cite{Bellhope,Bohmhope}.  As far as I know, neither commented on the ``two-state" or ``transactional" interpretations discussed below.  

An idea related to an absolute universal frame is a preferred frame which is identified with some specific  aspect of the physical environment.  Suggestions have been made \cite{CohenHiley} that the preferred frame is that of the cosmic microwave background, or  the experimental setup in the EPR-Bohm experiments, for instance its center of mass.  

The idea of a preferred frame attached to the center of mass of the experimental setup does not involve an absolute universal frame or time.  Hence it might be thought that it does not conflict with relativity.  However, it is questionable whether the notion of any sort of preferred frame that depends on the state of motion of the system is really in the spirit of relativity.  The idea is for the influence of a measurement on one subsystem to act instantaneously on the other in the ``time" of the center of mass frame.  Invoking this instantaneity seems to imply that the preferred ``time" has some sort of special significance {\it as} time, i.e. time of action -- yet it is tied to the center of mass of a particular arrangement.  Another way of putting it is that the relativistic symmetry of all inertial frames remains in this picture, but the physical significance of this symmetry becomes unclear.  In short, there is an amalgamation of relativistic space-time and a preferred frame, in which neither has a very clear meaning.   Much the same criticism can be made of proposals to tie the preferred ``time" to a universal frame, such as the cosmic microwave background.       

\subsection{Relativistic symmetry with ``retroaction"}

A second class of ideas for a solution to the conflict tries to maintain the relativistic symmetry.  Then the problem is how the correlations come about in a relativistically symmetric way.  The proposed solutions involve what might be called ``retroaction".  Essentially, each particle has communication of its own future with its own past.  One idea along these lines is the ``two-state" formalism \cite{twostate,Costa} of Aharanov and Vaidman and of Costa de Beuaregard, in which the idea of a quantum mechanical state is expanded to include a state propagating backwards in time, starting from the measurement, and going back to creation of the entangled pair.   Another idea is Nelson's ``transactional" interpretation of quantum mechanics \cite{transactional}.  This involves a kind of quantum mechanical utilization of the Wheeler-Feynman ``absorber" idea \cite{WF1,WF2}, in which there is ingoing information from a system's future as well as outgoing information from its past.      

The basic problem I see with these ``retroaction" solutions is that the past behavior of a subsystem is determined by its future interaction with a detection system, which may not even be set until the future.  That is, the past behavior of a subsystem is determined by {\it its own} future behavior, even though the future behavior supposedly is not yet determined!  This seems absurd on the face of it -- unless everything, including the future setting of the detector, actually {\it is} predetermined.    

Predetermination might appear to be compatible with conventional notions of causality combined with time reversal symmetry.  But, it would actually mark a drastic departure from conventional causality:  the future state of a system and its subsystems (the two entangled electrons), including its future interactions with {\it external} systems, would be completely reflected {\it in the state of the system alone}.   While this is logically possible, it flies in the face of my sense of how the world works.

Not the least reason is that it seems to preclude the possibility of free will.  If the state of a subsystem is already determined by its interaction with the detector, which I have not yet even set, and the explanation is that my future setting of the detector is pre-determined, then it must be the case that I have no freedom in setting the detector.  This seems unacceptable, at least with our present knowledge of the mind and the question of freedom of the will.  

Bell considered \cite{superdet}, the possibility that everything is ``superdetermined" -- a kind of universal predetermination.  Bohm and Hiley \cite{BohmHileyuft} also seem to have considered this kind of notion in the context Einstein's ideas about a generalized field theory.  This might be expressed in a ``retroaction" picture in terms of a universal wave function.  Bell considered the idea of superdetermination to offer logically a solution for the conflict of relativity and quantum mechanics, but nonetheless rejected it, at least partly on the grounds that it flies in the face of freedom \cite{superdet}.  Instead, he argued for ``a real causal sequence which is defined in the ether" \cite{realcausseq}, apparently being willing to pay the price of a return to the pre-Einsteinian notion of absolute time.

\section{The symmetric view: relativistic mutual influence} \label{symview}

In the proposal here, called the ``symmetric view", the aim is to deal with the relativistic asymmetry of present quantum mechanics in a way that (1) maintains Einsteinian relativity and eliminates the asymmetry in the present quantum description;  (2) maintains the empirical predictions of quantum mechanics, at least those which so far have been upheld in experiment; and (3) modifies conventional notions of causality in as mild a way as possible.  In particular, we want to avoid the problem of retroaction, and also the problem described later of mutual self-causation.     

In the symmetric view, the measurement process on entangled subsystems is modified in the way illustrated in Fig.~\ref{symview}, which is meant to be contrasted with the picture of the standard quantum mechanical description in Fig.~\ref{standardview}.  It is proposed that measurements on entangled subsystems are connected by a reciprocal mutual influence, for systems which are space-like separated.  More generally, ``measurement interactions" of the subsystems with the environment, i.e. interactions in which measurable information is carried away, are connected by the symmetrical, mutual influence; I make no distinction between such an interaction and a measurement.  Because of this reciprocity, there is complete symmetry of the description of the events of both subsystems.  Einsteinian relativity of simultaneity and space-time symmetry are completely preserved.   The events at A and B and their correlations are the result of the mutual influence between 1 and 2 across space-time.   

Needless to say, this mutuality requires faster-than-light influence between space-like separated events.  The two-way mutual influence takes place across time in all frames, with backward-in-time transmission in one direction, and forward in time in the other direction, except the unique frame in which both measurement interactions are instantaneous.  I call this special frame the ``interaction frame", about which there will be more to say later.  

The symmetric view with mutual influence evidently must involve a modification of conventional notions of causality and of temporal sequence in causation.  What can we say about this?

\section{What is the nature of the mutual influence?}

If one accepts that the symmetric view removes the conflict with relativity, there nonetheless are mysterious things about the symmetric mutual influence and how it is conveyed.   I first consider the nature of the influence, and turn later to the question of the means by which it is conveyed.   Clearly, at this highly speculative stage these questions cannot be entirely separated.     

We are used to thinking that ``an event at A causes an event at B".    Usually, the understanding is that event A is earlier than or simultaneous with event B, as in the ordinary quantum description.  However, the question is deeper than that posed simply by a later event causing an earlier event.    Are we in danger in the symmetric view of saying that the outcome of an event at A causes the outcome of an event at B {\it and} that the outcome of the event at B causes the outcome of the event at A, as if A and B were each the cause of its own cause?  It turns out that while usual notions about temporality and time sequence must be modified, the much more drastic change to a conception of ``mutual self-causation" can be avoided.   There are several possibilities for how this could work. 

One possibility is a kind of mutual ``passive" influence.  The idea is that each measurement event only needs to ``know" what is being measured, i.e. what direction of spin, at the other event.  Then the influence involves each subsystem (A) merely ``telling" what is being measured on it  (or what is happening in the measurement interaction, see remarks shortly on contextuality) to the other subsystem (B), rather than acting on the other subsystem  in the sense of A ``telling B what to do" on the basis of a perhaps yet-to-be determined {\it outcome} of the measurement at A.  It turns outs that this passive influence is all that is necessary to get the Bell correlations.    

How this could work is perhaps seen most readily in a ``realistic" picture.   This can be described in colloquial terms of  a ``program" associated with each subsystem, which tells the subsystem how to react to a given measurement.   This could be consistent with a kind of hidden variables picture, but at this stage can just as well be thought of as simply a useful metaphor.  (The term ``hidden variables" to some has far more restrictive connotations than the idea that systems might have ``real properties", unknown to us at present, that underlie their quantum behavior.   The distinction, which seems to have been very important to Einstein, is discussed forcefully by Fine \cite{shaky}).       

First, consider a realistic picture in which the Bell correlations are {\it absent}.  Initially, each particle in the entangled state is programmed to react with a certain result, i.e. + or $-$, when subjected to a measurement in a given, arbitrary direction.   Thus, particle 1 is programmed to react to a measurement $\mu$ with outcome + or $-$.  Particle 2 is programmed to react to a measurement $\nu$ with outcome + or $-$, subject to the constraint that the programs of 1 and 2 be consistent.  That is, if $\mu, \nu$ are the same, particle 2 is programmed $-$ if particle 1 is +, and vice-versa.

This picture can be entirely consistent for the more complex kinds of measurements where the theorems of Gleason \cite{Gleason}, Kochen and Sprecker \cite{KochenSprecker}, and Bell (his ``first theorem" \cite{Bellsfirst}) come into play \cite{Mermin2th}.  That is, the program could be contextual, such that the response of a subsystem is determined by its real properties, but only given the context of the measurement being performed locally on the subsystem.  Furthermore, entirely consistent with contextuality, the program could involve the real properties not just of the subsystem, but subsystem and measurement system interacting together, if one favors as I do the avoidance of any distinction between ``measurement" and ``measurement interaction".      

The contextual programs could operate {\it independently} for the separated subsystems of an entangled system, given the initial local interaction at their formation, except for one thing:  the Bell correlations are missing -- this could be said to be the whole point of ``Bell's theorem" (his ``later" theorem \cite{Bellstheorem})!  The reason is that so far everything  that happens is localized in space -- formation of the entangled system, and subsequent measurements on the subsystems -- and the Bell correlations cannot have been built into the programs.

However, the Bell correlations can arise if 1, 2 re-program to react properly as the complete context demands, including events at separated subsystems.  All that is necessary is that each particle receive information of what property (spin direction $\mu, \nu$) is being measured on the other, (or alternately, information of the real properties of the other subsystem interacting with measurement system).  The Bell correlations can result if 1 and 2 each re-program in light of knowledge of the measurement being performed  at the other, in such a way that an ensemble of such measurements $\mu, \nu$ yields the proper statistical results.  

Another possibility, perhaps only superficially different, is that the mutual influence involves not simply an exchange of information of the measurement process at the other subsystem, but rather some sort of ``active" mutual dynamical interaction.   For example, in a realistic picture, the programs might involve a sort of generalized Bohmian quantum potential, in which A and B mutually determine the quantum potential, which acts on jointly on both.  This notion will be revisited after further discussion of the ``interaction frame".  
 
It must be emphasized  that the symmetric view need not require an underlying realistic picture -- though I find such a picture very appealing.  The programs described above could all be orthodox quantum mechanical ``dice-throwing", or might involve some sort of stochastic hidden variables.  The difference from either orthodox quantum mechanics or the usual stochastic hidden variables is that in these pictures, a measurement on one subsystem instantaneously ``reloads the dice" for the other; whereas in the symmetric view, the reloading of the dice happens symmetrically via a mutual influence between the two subsystems.

\section{how is the influence conveyed?}

The symmetric view removes the asymmetry of present quantum mechanical description of entangled systems by introducing a mutual, two-way influence which is completely symmetric in all reference frames.  This raises the obvious question of the mechanism or means by which the mutual influence is conveyed, to which we now turn.    

The problem is that for A to convey its influence to B, and conversely for B to convey its influence to A, each must deliver its information to the right destination in space-time, that is, to the other measurement event.   Does each measurement event ``know" where the other is?   Either answer leads to perplexities of its own.

\subsection{Enframing by the interaction frame}

One logical possibility is that each measurement event knows where the other is in space-time when it conveys its influence.  How could this be? 

An evident point about the mutual two-way influence, referred to already, is that it is instantaneous in the unique frame -- earlier called the interaction frame -- in which the measurement events are simultaneous.    The idea is that the interaction frame constitutes a sort of ``channel" between A and B by which the influence is conveyed.  This means that there is an {\it active role for space-time and its frames}.  A way of putting it is that there is an ``enframing" of the measurement events by the frame in which they are instantaneous.  An image of this enframing is that each inertial frame actively searches along all of its space-like directions for entangled measurement interactions.   When one of the frames finds entangled measurements -- the unique frame in which the measurements are simultaneous -- it establishes the channel through which the mutual influence is exchanged, as illustrated in Fig.~\ref{enframing}.  The interaction frame thereby takes on an active physical role, but this does not in any way involve a loss of relativistic symmetry.  If anything like this happens, it constitutes a salient extension of the significance of the notion of frame beyond its purely conventional significance in classical special relativity.  

An obvious question is what happens to this picture in curved space-time.  A very natural possibility is that the influence is conveyed along geodesics connecting the two sub-systems; this remark applies as well to the scenarios considered in the next two subsections.

\subsection{Broadcasting and reception}

Another possibility is that A and B {\it don't} know where each other are.  One way this might work is that each measurement event ``broadcasts" its influence without knowing where in space-time the other is.  The most natural possibility would be that each event broadcasts in this non-classical way through the space-like quadrants of its own light cone.   This would give coverage for every possible reference frame for which there is a problem, but limit the broadcasting to space-like separated events.  

The obvious problem with this is how to avoid problems with causal loops.  If there are causal loops, this leads to the same problems of retroaction, etc. which we have said plague other proposed solutions to the conflict of relativity and quantum mechanics.  

Perhaps this is not the end of the story.  One can imagine a broadcast influence which can only be {\it received} by the other interaction event, as illustrated in Fig.~\ref{broadcastreceive}.   This is symmetric in a way somewhat like quantum field theory.  The influence is analogous to a particle that can only be emitted by one interaction event and absorbed by the other.  If the influence is not absorbed, no information is conveyed, and the problem of causal loops is avoided.

This may not be so much different from the idea of the interaction frame.  In the interaction frame picture, 1 and 2 find out where their partner is by means of an active search of both measurement events by all frames, with ``enframing" by the unique frame of simultaneity and establishment of a channel .  This may not be so much different from the  picture of an influence which is broadcast in all frames, but only effective in the unique frame, again, the interaction frame, in which there is both a broadcaster and receptor in each direction.

\subsection{An active dynamical interaction?}

This is the point to return to the possibility the mutual influence involves not simply an exchange of information, but rather some sort of ``active" mutual dynamical interaction.  An example might be an adaptation to the interaction frame  of the picture given by the Bohmian potential.    The standard Bohmian picture is discussed above and in detail by Bohm and Hiley \cite{undivided} and Albert \cite{Albert}. The Bohmian picture is an example of contextual hidden variables; the measured outcomes of experiments are not pre-existing real properties of the particles themselves.  Which particle is determined to have spin + or $-$ when spin in the $z$ direction is measured on both particles is determined by which particle first enters a detector, in an unspecified preferred frame.   The other particle then is influenced by a very rapid change in the quantum potential, associated with a rapid change in the wave function of the particles and measurement apparata (there is no collapse).  In response to the change in the quantum potential, the second particle rapidly ``swerves" in the opposite direction to the first, so one particle is measured + and the other $-$.   Which particle is measured + and which $-$ depends on which enters a detector first, which of course is grossly non-relativistic. 

In the symmetric view, this picture would have to be modified.   In the frame in which both particles enter  the measurement interaction symmetrically and simultaneously, i.e. the interaction frame, there is no temporal asymmetry in the evolution of the quantum potential to determine which particle is measured + and which $-$.  Instead, the asymmetry would come from the relative position of the particles with respect to the measurement devices.  

This sketch of a ``symmetric Bohmian picture" is given merely as an example of how an active dynamical interaction might be possible in a realistic description in the symmetric view.  It is an example of how the experimental outcome, consistent with Bell inequalities, might be obtained by a mutual exchange of information about both the particles and the measurement system -- the information in this example consisting of the evolving quantum potential.

\section{Is an empirical test possible?}

Our starting point was the observation that the standard quantum mechanical description seems to imply a preferred or absolute space-time frame in which the influence between entangled subsystems is transmitted; yet any space-time frame appears to be as good as any other for the role of preferred frame, since the predicted correlations are independent of the choice of frame.   

The symmetric view advocated here is that there is no reason to posit a preferred frame, independent of the measurement events; rather, the correlations arise from a symmetric mutual influence between the measurement events.  

Is it conceivable that there could be an empirical test among these pictures  of how the Bell correlations are conveyed, or for the various hypothesized frames (absolute frame, interaction frame, system-dependent special frame)?   One way might be if the quantum mechanical influence could be {\it blocked} in the hypothesized frame in which the correlations are carried, i.e., if something could intervene to prevent or modify the correlations.   In this connection it is interesting that ``frame jamming" has been considered  \cite{jamming}.  Under certain conditions, it would be possible to distinguish empirically between either the symmetric view or the view that there is some kind of absolute frame.       

How this would work is illustrated in Fig.~\ref{frameblocking}, which considers two experiments.   Panels (a) and (b) sketch two possible outcomes of the first experiment.  Panel (a) supposes that the Bell correlations occur via an influence conveyed according to the usual quantum description, i.e., instantaneously in the preferred frame.  If a jamming device is interposed in this frame, the influence is blocked and the correlations are missing or altered.   Panel (b) shows the same experimental setup but under the supposition that the correlations occur as in the symmetric view.  In this case the jammer  does not block the influence, and the correlations are present.      

On the other hand, consider a second experimental setup in which the jammer is moved as envisioned in panels (c) and (d).  Panel (c) supposes the standard view, and panel (d) the symmetric view.  Now, the correlations will be present if the standard view (c) is valid, and absent if the symmetric view (d) is valid.

If the two experiments sketched in panels (a - d) could be realized, together they would give evidence for the validity of either the standard or symmetric view.  If the standard view is correct, the correlations would be absent in experiment 1 and present in experiment 2.  If the symmetric view is correct, the correlations would be present in experiment 1 and absent in experiment 2.  

Of course, the realization of these experiments depends on very speculative assumptions.  Not only is it necessary to be able to construct a jamming device, but the jammer must block only one of the two alternative means of conveyance of the influence.  If the jammer is such that it blocks both alternatives, the effect would be to jam the correlations in both experiments.  Then no empirical discrimination between the standard and symmetric views would be possible.

\section {Epilogue:  Bell, Einstein, Bohr, and Causality}

Einstein famously said that ``God does not play dice".  He also famously said that he was bothered by the ``spooky action at a distance" that he saw in the situation envisioned in the EPR thought experiment.  Einstein's position is set out very lucidly in his Autobiographical Notes \cite{Einsteinnotes,philsci}.  He thought the particles electrons must have either (1) pre-existing outcomes determined by ``real properties" of the system  (no dice-playing), or (2) an effect of the first measurement on the ``real situation" of the second via ``spooky action at a distance", or as he put it in his Notes, ``telepathically".   He rejected the second possibility as ``entirely unacceptable".  Indeed, Einstein rejected classical action at a distance precisely on the grounds that ``there is no such thing as simultaneity of distant events" \cite{Einsteinnotes61}, i.e. that it lacks relativistic symmetry.  

Einstein did not live long enough to learn about the Bell correlations and the extraordinary way in which they bring out the extraordinary difficulty of reconciling quantum mechanics and relativity.  I think it entirely possible that he foresaw, but rejected as implausible, something like the Bell correlations (without knowing their exact nature as predicted by quantum mechanics).    The predictions of Bell (and their empirical verification) show that possibility (1) is untenable:  the outcomes cannot be due to pre-existing real properties.  It seems that there must indeed be ``spooky action at a distance".   Mermin \cite{misfortune} called it a ``misfortune of intellectual history" that Einstein did not live long enough for us to know his reaction.

Bell evidently was immensely disturbed at the implications of his theorem, which he apparently was inspired to investigate largely from pondering Einstein's critique of quantum mechanics \cite{BellBernsteinreason}.   His response to his discovery was a call for the correlations to be explained in terms of ``a real causal sequence which is defined in the ether" \cite{realcausseq} .  He said \cite{BellBernsteinreason} ``For me, it is so reasonable to assume that the photons in those experiments carry with them programs, telling them how to behave."  In this, Bell seemed to be following Einstein's call for a description in terms of pre-existing real properties \cite{Einsteinnotes}.  In light of his correlations, Bell said that ``it is a pity that $\cdots$ the reasonable thing just doesn't work".   

In view of the empirical facts the next most reasonable thing for Bell was to suppose that the correlations came into being through a causal interaction, that of the measurement on one particle acting on the other -- a ``real causal sequence".  Because it is a {\it sequence}, Bell was willing to go so far as to abandon relativity and return to absolute time.  The other possibility Bell saw was that everything is superdetermined \cite{superdet}, as described above.  Bell saw that the price for this is that there is no freedom.  This price he was very reluctant to pay.  On the other hand,  Einstein said that he did not believe in freedom of the will \cite{Einstein_nofreewill}.  Interestingly, Bohm and Hiley considered an explanation of the Bell correlations along the lines of a superdetermination connected with Einstein's ideas for a unified field theory \cite{BohmHileyuft}.

It is sometimes said that Einstein's conceptions of space and time are not ``revolutionary".  The facts of the Bell correlations revealed by quantum mechanics show how revolutionary they really are!  Classical relativity veils the revolutionary character of relativistic space-time, because it avoids the conflict with ordinary conceptions of causality, which are perhaps equally as basic as concepts of space and time.  Quantum mechanics forces us to decide whether to take the relativistic conception of space and time seriously.   If we do so, it seems to require a significant modification of ideas of causation.    

The symmetric view is consistent with realism and relativistic symmetry.   It involves a mutuality of causation as a  feature of physical description.  How this could operate, e.g. via ``enframing by the interaction frame",  is at this point only a matter of speculation.  But it seems to me the viewpoint most consistent with the empirical facts, relativistic symmetry, and reasonable notions of causality.  Whether or not it entails ``spooky action at a distance" and ``telepathy", if it should turn out to have anything to do with physical reality, it will be an example of the use of symmetry as a heuristic principle to uncover features of the world -- an approach that Einstein certainly put to good use.      

As a final remark, it is interesting to consider the status of causality in the symmetric view in relation to Bohr's response \cite{bohrepr,bohrnotes} to the EPR paper \cite{epr} and to Einstein's statement of his views in his Autobiographical Notes \cite{Einsteinnotes}.  Bohr advocated \cite{bohrnotes238} that the word ``phenomenon" refer to ``an account of the whole experimental arrangement".  If by this Bohr had in mind the {\it unity} of the phenomenon, I  find this largely consonant with the symmetric view, especially the idea of the interaction frame.  There is a certain irony here, in that the motivation for the symmetric view is to preserve Einsteinian relativity.  (Bohr's view that ``a more detailed analysis $\cdots$ is {\it in principle} [Bohr's italics] excluded" \cite{bohrnotes235} runs very counter to a description in terms of ``real properties" such as I find desirable.)  In contrast, the position of Bell and Bohm that one should go back to an absolute time with ``real causal sequence" seems much less consonant with the notion of the unity of the phenomenon.


\begin{figure} \caption  {Standard view of the measurement process on entangled subsystems.  A one-way influence is conveyed in a preferred frame by the measurement event which is temporally first in that frame. \label{standardview}}\end{figure}

\begin{figure} \caption{Symmetric view of the measurement process on entangled subsystems. A symmetric two-way mutual influence is conveyed between the two measurement events.  \label{symview}}\end{figure}

\begin{figure} \caption{Enframing by the interaction frame.  Each frame searches for entangled measurement events; the frame in which they are simultaneous is the interaction frame, which builds a quantum channel between the  subsystems.\label{enframing}}\end{figure}

\begin{figure} \caption{Broadcasting and reception.  Each measurement event broadcasts information which can be received only by the other measurement event, setting up a channel for the symmetric mutual influence.  \label{broadcastreceive}}\end{figure}

\begin{figure} \caption{Experiments to test between the standard description and the symmetric view.  In (a) and (c), the standard view is assumed to hold; in (b) and (d), the symmetric view is assumed to hold.  \label{frameblocking}}\end{figure}

\end{document}